# Towards Quantitative Interpretation of 3D Atomic Force Microscopy at Solid–Liquid Interfaces

Qian Ai[1,2], Lalith Krishna Samanth Bonagiri[2,3], Amir Farokh Payam[4], Narayana R. Aluru[5], Yingjie Zhang[1,2,6]*

1. Department of Materials Science and Engineering, University of Illinois, Urbana, Illinois 61801, United States

2. Materials Research Laboratory, University of Illinois, Urbana, Illinois 61801, United States

3. Department of Mechanical Science and Engineering, University of Illinois, Urbana, Illinois 61801, United States

4. Nanotechnology and Integrated Bioengineering Centre (NIBEC), School of Engineering, Ulster University, Belfast BT15 1AP, United Kingdom

5. Walker Department of Mechanical Engineering and Oden Institute for Computational Engineering & Sciences, The University of Texas at Austin, Austin, Texas 78712, United States

6. Beckman Institute for Advanced Science and Technology, University of Illinois, Urbana, Illinois 61801, United States

*Correspondence to: yjz@illinois.edu

**Abstract:** Three-dimensional atomic force microscopy (3D-AFM) has been a powerful tool to probe the atomic-scale structure of solid–liquid interfaces. As a nanoprobe moves along the 3D volume of interfacial liquid, the probe–sample interaction force is sensed and mapped, providing information on not only the solid morphology, but also the liquid density distribution. To date 3D-AFM force maps of a diverse set of solid–liquid interfaces have been recorded, revealing remarkable force oscillations that are typically attributed to solvation layers or electrical double layers. However, despite the high resolution down to sub-angstrom level, quantitative interpretation of the 3D force maps has been an outstanding challenge. Here we will review the technical details of 3D-AFM and the existing approaches for quantitative data interpretation. Based on evidences in recent literature, we conclude that the perturbation-induced AFM force paradoxically represents the intrinsic, unperturbed liquid density profile. We will further discuss how the oscillatory force profiles can be attributed to the probe-modulation of the liquid configurational entropy, and how the quantitative, atomic-scale liquid density distribution can be derived from the force maps.

**Table of Contents Graphic**

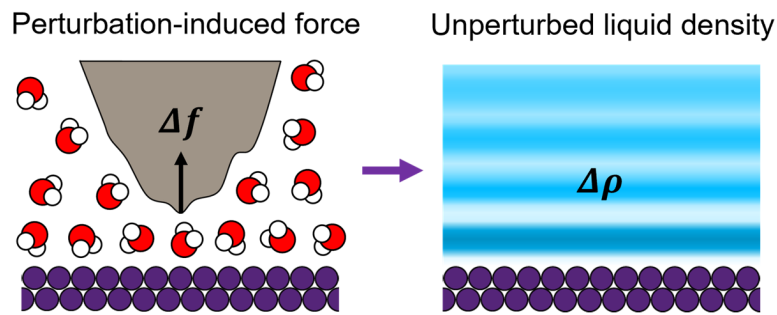

# 1. Introduction

Solid–liquid interfaces are ubiquitous in various natural and engineered systems, and are critical for a large variety of functions, such as biomolecular signal transduction,[1–3] water filtration,[4–7] corrosion control,[8–10] and electrochemical energy conversion and storage.[11–18] At the transition region between the solid surface and the bulk liquid, the interfacial molecules form solvation layers, also called the electric double layers (EDLs) if charged species are present. Such interfacial liquids are oftentimes as important, if not more, as the solid surfaces in modulating the chemical, electrochemical, and biological functions.[1,11,19] However, to date our understanding of the interfacial liquid structure is still highly limited, despite all the advanced computational and experimental tools developed in the past few decades to unravel the structure of matter. Computationally, the highly complexity of mobile interfacial liquid species poses significant challenges for ab initio approaches;[20,21] meanwhile, classical molecular dynamics (MD) simulations are constrained by the accuracy of force fields in capturing realistic intermolecular and liquid–solid interactions.[21,22] Experimentally, the buried nature, dynamic state, and volatility of interfacial liquids also present significant challenges for their characterization.[23–25]

Due to the difficulty in probing the interfacial liquid structures, oftentimes they have been treated as continuum dielectric environments for simplicity, either in computation (i.e., implicit solvent model) or practical device/systems design (e.g., supercapacitors, fuel cells, batteries, etc.).[26,27] However, these continuum models have been insufficient to explain many experimental observations, such as the anomalously low dielectric constant of interfacial/confined water,[28] the unconventional double layer capacitance of highly ionic electrolytes,[29] and the significant electrolyte-dependence of many electrocatalytic reactions.[30–32] These emergent interface-related "anomalies" provoke a key question: are we perceiving interfacial liquid structures properly, even to the first order approximation? Specifically, what is the spatial distribution of the interfacial liquid species, in the "simple" equilibrium state after time-averaging? Besides the charged species that are known to accumulate or deplete in response to interfacial electric fields, do the neutral species simply have uniform density everywhere?

Among the existing experimental methods, only two types of techniques can provide information on the spatial density profiles of interfacial liquids, to the best of our knowledge. One is X-ray/neutron scattering, which enables the extraction of electron/mass density distribution along the z direction (normal to the solid surface).[33–36] These scattering techniques are only limited to flat, crystalline solid–liquid interfaces. In the presence of heterogeneities, such as vacancies, impurities, step edges, and random roughness, it is extremely challenging, if at all possible, to extract spatial liquid density information near the heterogeneous sites from scattering data. Another method is 3D-AFM, which measures the 3D atomic-scale force maps between the AFM probe and species surrounding the probe at the solid–liquid interface.[37–52] This real-space imaging technique is sensitive to both the solid surface and the nearby liquid, and can detect both the flat interfaces and locally heterogeneous structures. As to the quantification of liquid structure, it has been generally accepted that the 3D force maps are closely related to the liquid density distribution at the interface.[25] However, the exact quantitative force–density relation as well as the possibility of probe-perturbation effects remain under debate.

In this perspective, we begin by discussing the technical details and force mapping algorithms of different modes of 3D-AFM. Next, we review the current understanding of intrinsic interfacial liquid distributions. We will then survey the existing analytical and computational methods used to link 3D-AFM force maps to liquid density profiles. In particular, we will highlight the solvent tip approximation (STA),[53,54] where the tip of the AFM probe is modeled as a single solvent molecule, whose interaction with the nearby liquid is regarded as the approximate tip–liquid interaction. We will further discuss the thermodynamic origins of the probe–interface interactions, and explain why the experimentally measured force contains information on the intrinsic liquid density profiles in the absence of the probe. Finally, we will rationalize why STA is a surprisingly accurate approximation for quantitatively determining interfacial liquid density profiles. We will also discuss future directions for 3D-AFM data quantification and its applications.

**2. Operation Mechanisms and Key Results of 3D-AFM**

The overall 3D-AFM setup and imaging process is shown in Figure 1. An AFM probe is immersed in bulk liquid and brought to the solid–liquid interface (Figure 1a). The liquid is oftentimes in the form of a droplet spread on a solid surface, except the setup demonstrated in our lab where the liquid is contained in a fully sealed electrochemical cell.[49–52] Common to all existing 3D-AFM methods, a 3D scanning algorithm was used to measure the force maps (Figure 1b). At room temperature, there is always a finite thermal drift of the tip and sample position. In order to obtain accurate 3D molecular images (e.g. in a $10 \times 10 \times 5$ nm$^3$ volume), the data acquisition needs to take place within ~1 min, requiring a z rate in the scale of ~100 Hz. At such high z rate, the traditional triangular waves used for force curves/maps can generate significant noise at turning points (lowest/highest z points).[55] To overcome this challenge, the 3D-AFM community has adopted sinusoidal waves to drive the tip motion in the z direction, while simultaneously moving the tip along x and y directions through linear scan.[25,42] This sinusoidal wave suppresses noise and enables a z rate up to 1 kHz, although the signal-to-noise ratio (SNR) is usually lower at higher z rate.

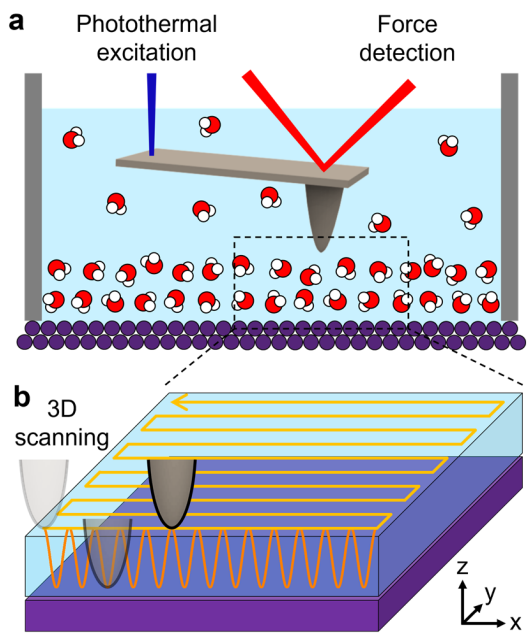

**Figure 1. 3D-AFM schematic.** (a) 3D-AFM setup, including the solid substrate, liquid/electrolyte, AFM probe immersed in liquid, photothermal excitation at the base of the microlever, and laser deflection-based force detection. (b) 3D scanning process to enable high-speed force mapping.

While the scanning algorithm is nearly the same for all the existing 3D-AFM measurements, the force detection mechanism has significant variations. Overall, the force measurements can be classified into two categories, DC (quasi-static, non-resonant) and AC (dynamic, resonant) mode. The AC mode can be achieved via either amplitude modulation (AM) or frequency modulation (FM). The detailed mechanisms of these different modes and the key findings on solid–liquid interfaces are summarized below.

**2.1. DC mode 3D-AFM.** In DC mode, the AFM cantilever is quasi-static, i.e., no additional oscillation besides the 3D scanning process. The z rate during 3D scanning (Figure 1b), up to 1 kHz, is much lower than the first eigenmode of the probe and thus cannot excite the resonance oscillation of the probe by itself. Force is measured directly via the laser deflection off the end of the cantilever, which can be achieved in most of the existing commercial AFMs. Before the development of DC mode 3D-AFM, DC force curve measurements at single (x, y) points of solid–liquid interfaces were already demonstrated by a few labs including Atkin, Balke, Andres, and Espinosa-Marzal.[56–59] These force curve measurements were mostly performed in the highly viscous ionic liquids which leads to pronounced force spikes due to the so-called "molecular layering". However, due to a lack of in-plane spatial resolution, the quality/cleanliness of the solid surface was oftentimes not known, and the point-based force curves tend to have large fluctuations, making it challenging to perform statistically significant analysis.

Our lab demonstrated DC mode 3D-AFM in the past few years, and used it to obtain 3D force maps of ionic liquids and water-in-salt electrolytes (Figure 2).[49–51] With sufficiently low mechanical noise, we achieved lattice resolution of graphite and $MoS_2$ surfaces, proving the atomic-level cleanliness of the substrate. The 3D force maps at clean substrate areas are highly consistent among different batches of samples/measurements, enabling statistically significant analysis of the interfacial liquid profiles. From these results, we observed persistent force oscillations where the interlayer distance beyond the first peak is nearly constant for a given electrolyte. At different electrode potentials or substrates, we only observed changes in the first layer due to electrostatic modulations, while upper layers remained mostly unchanged (Figure 2c, f). These results indicate that the upper layer configuration is likely primarily influenced by the bulk intermolecular structure of the liquid, while only the first layer depends sensitively on the interaction with the substrate.

While the DC mode 3D-AFM provides a convenient method to probe the interfacial liquid profiles and is broadly applicable in the high viscosity limit (e.g., viscosity larger than ~20 cP), it is usually not sensitive to interfacial density variations for less viscous liquids (e.g, viscosity lower than ~10 cP) due to the lower force oscillations of these systems. In addition, due to the low spring constant of the cantilever (typically less than ~10 N/m to ensure sufficient sensitivity of the deflection/force), the probe can have "snap-through" transitions as it goes through each liquid layer.[49–51] This non-linear, out-of-equilibrium behavior makes it highly challenging to quantitatively convert the force maps to liquid density distributions.

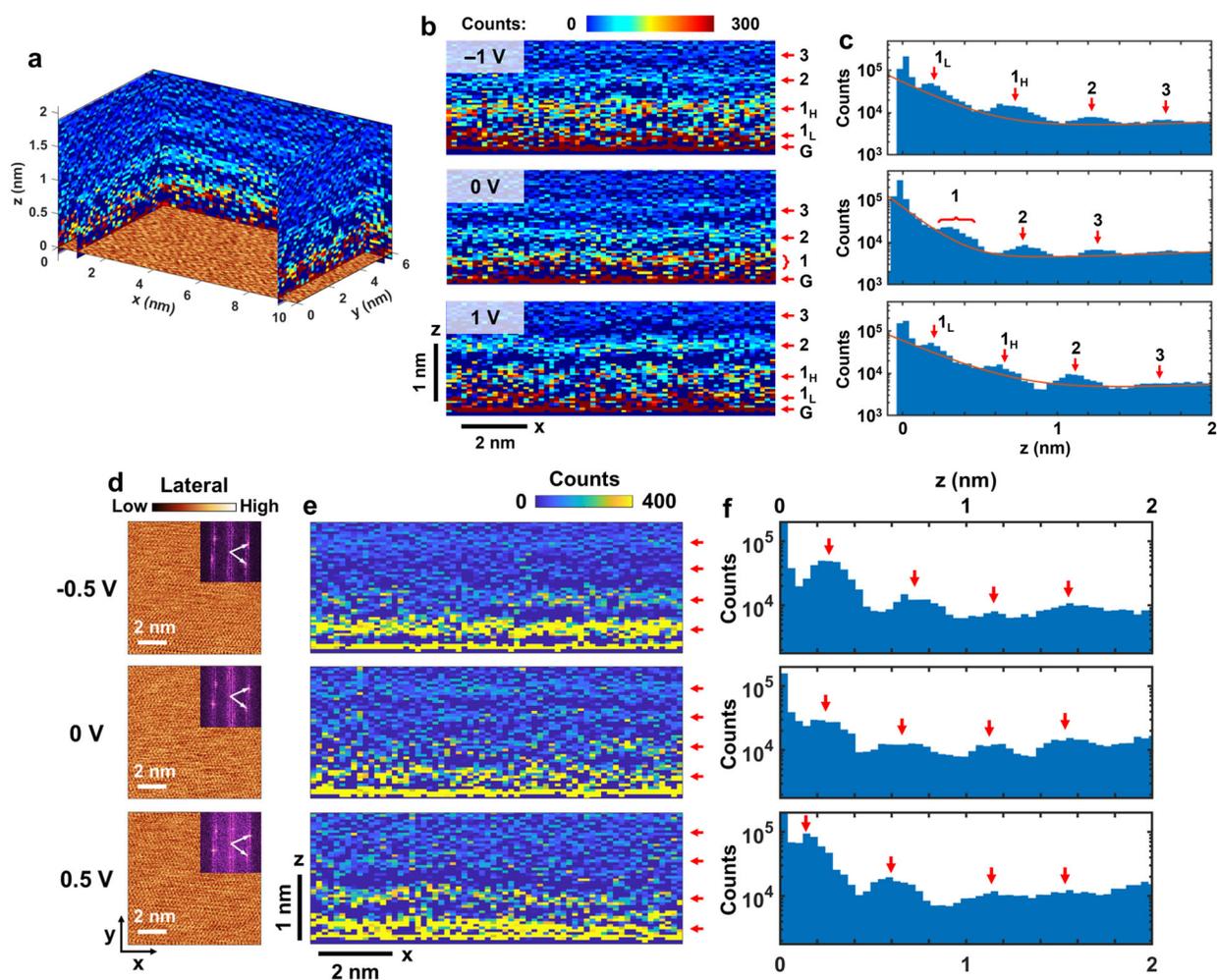

**Figure 2. Examples of DC mode 3D-AFM results.** (a–c) 3D count maps, x-z cross sections, and z count histograms of 1-ethyl-3-methylimidazolium bis(trifluoromethylsulfonyl)imide (EMIM-TFSI) on HOPG.[49] Reproduced with permission from ref [49]. Copyright 2020 American Chemical Society. (d–f) Substrate lateral force maps, x-z count maps, and z count histograms of HOPG/21 m LiTFSI in water interface.[51] Reproduced with permission from ref [51]. Copyright 2022 American Chemical Society.

**2.2. AC mode 3D-AFM.** In AC mode, the AFM cantilever is excited at resonance frequency, at either the first or higher eigenmode. Excitation is usually achieved through photothermal effects, where an intensity-modulated laser focused near the base of the microlever induces periodic bending and oscillation of the lever (Figure 1a). This photothermal method selectively excites the microlever without inducing mechanical oscillation of the support chip and probe holder, thus avoiding the large mechanical noise and "forest of peaks" problems that are known as the key bottlenecks for the traditional piezoacoustic excitation method in liquid.[60–62]

A resonantly-excited cantilever can be modeled as a driven damped harmonic oscillator, and the probe–sample interaction results in a change of the resonance frequency, amplitude, and/or phase of the oscillator. In the FM mode, resonance frequency shift ($\Delta\omega$) is measured, while in AM the amplitude ($A$) and phase shift ($\phi$) are obtained. Using the harmonic oscillator model, $\Delta\omega$ (FM mode) and $A$ and $\phi$ (AM mode) can both be converted to the total conservative force ($f$) with reasonable accuracy.[63,64] The force typically consists of two components: a repulsive background that resembles an exponential function and an oscillatory decay signal ($\Delta f$).[37,38] In existing articles on 3D-AFM, $f$ or $\Delta f$ maps were oftentimes reported as the "structure" of interfacial liquids.[37,38,45,65]

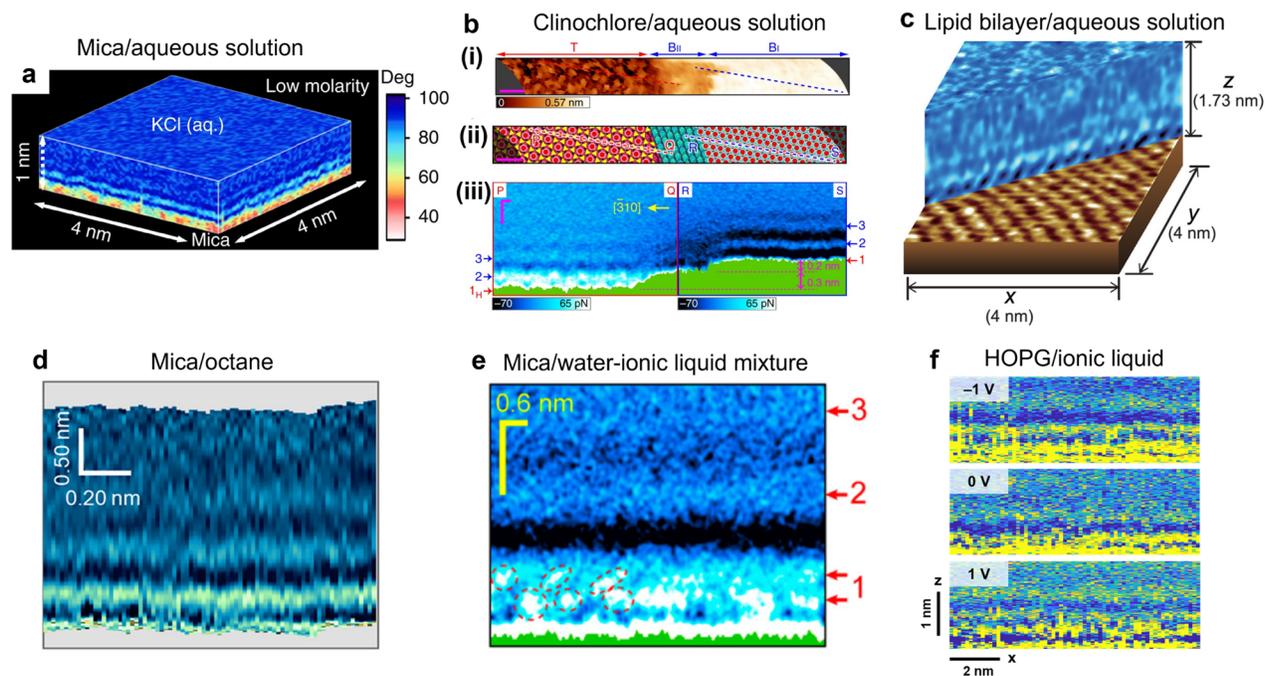

**Figure 3. Gallery of AC mode 3D-AFM images of solid–liquid interfaces.** (a) AM 3D phase map of 0.2 M KCl in water on mica.[39] Reproduced from ref [39]. Copyright 2016 The Authors. (b) FM maps of 0.1 M KCl in water on clinochlore surface, including (i) x-y topographic image, (ii) surface structural model, and (iii) x-z force map of a local area.[46] Scale bars: 1 nm (i, ii) and 0.3 nm (iii). Reproduced from ref [46]. Copyright 2017 The Authors. (c) FM 3D frequency shift map of 4-(2-hydroxyethyl)-1-piperazineethanesulfonic acid (HEPES) buffer solution (10 mM HEPES, 100 mM sodium chloride in water, pH 7.4) on a lipid bilayer.[65] Reproduced with permission from ref [65]. Copyright 2012 American Chemical Society. (d) AM x-z force map of octane/mica interface.[38] Reproduced with permission from ref [38]. Copyright 2020 American Chemical Society. (e) FM x-z force map of 30 mol% 1-ethyl-3-methylimidazolium dicyanamide (emim-DCA) in water on mica.[47] Reproduced with permission from ref [47]. Copyright 2020 American Chemical Society. (f) Potential-dependent AM x-z phase maps of EMIM-TFSI on HOPG, where the potential is against a Pt quasi-reference electrode.[49] Reproduced with permission from ref [49]. Copyright 2020 American Chemical Society.

FM 3D-AFM has been implemented by a few groups worldwide including Fukuma,[42,44,66] Yamada,[46,47] and Kühnle,[43,67] while AM 3D-AFM has been achieved by Garcia,[37–39,41] De Yoreo,[48,68] and our lab.[49,52] Overall, the AC mode 3D-AFM has been used to image a diverse set of solid–liquid interfaces, where the liquid includes pure water, aqueous solutions, organic solvents, ionic liquids, etc., while the solid includes mica, calcite, highly oriented pyrolytic graphite (HOPG), lipid bilayers, etc.[38–49,65–68] Figure 3 shows examples of these 3D-AFM images, where overall similar SNR was achieved comparing AM and FM modes. Despite the distinct liquid composition and solid structures, all of the reported 3D-AFM images reveal multiple layers (i.e., force oscillations), to the best of our knowledge. $\Delta f$ usually exhibits damped oscillations vs z, and the decay length is typically close to the interlayer distance. The number of layers/oscillation periods that can be observed depends on the oscillation strength of the interfacial liquid vs the force sensitivity of the measurement. In most cases, the oscillation becomes weak/negligible after 2–3 layers. Note that, to reliably sense the interfacial liquid structure, the free oscillation amplitude of the cantilever needs to be sufficiently small, at least lower than the interlayer distance of the imaged liquid.

A key prerequisite for AC mode 3D-AFM is the ability to excite resonance oscillation of the cantilever. In the viscous liquid environment, the quality factor (Q) of the probe tends to be reduced compared to that in air. If Q factor is larger than 1, the first eigenmode can usually be excited. In existing reports, liquids with viscosity between 0–40 cP have been successfully imaged by AC mode 3D-AFM.[38–49,65–68] In addition, due to the large spring constant (between ~10–100 N/m) of the probes used for AC mode, static cantilever bending is usually negligible in the 3D scanning process. Therefore, the oscillating probe is in a quasi-equilibrium state throughout the whole imaging process, and the conservative force sensed by the probe is independent of the z rate and is expected to be directly related to the local liquid density profile.

Since AC mode 3D-AFM is more broadly applicable and quantifiable than the DC mode, in the following sections we will focus exclusively on the AC mode measurements, with the aim of relating the experimentally determined conservative force to the intrinsic liquid density distribution.

## 3. Intrinsic Density Profiles of Interfacial Liquids

Before evaluating the quantitative connection between 3D-AFM force maps and the liquid distribution, it is important to know the "ground truth"—the intrinsic, unperturbed interfacial liquid structure. The precise atomic details of the interfacial liquids depend on the specific molecular interactions that can vary significantly among different liquids. To date many of such details are still under debate even for the most common liquids. Perhaps the most famous example is interfacial water, where the exact hydrogen bond configurations still cannot be precisely quantified through state-of-the-art computational and experimental approaches.[23,69] Here it is not our intention to unravel the most accurate atomistic interactions of all the relevant liquid systems. Rather, we focus on the common features of interfacial liquids that have been reproduced and verified by multiple theoretical/computational/experimental methods. A key structural feature we examine is the liquid/molecular density distribution, which directly regulates many crucial local processes such as capacitive charge storage, redox charge transfer, and mass transport.[26,70]

Two widely known and used theories describing the solid–liquid interfaces are the Gouy-Chapman-Stern (GCS) and Derjaguin–Landau–Verwey–Overbeek (DLVO) models.[26,27] GCS uses Poisson-Boltzmann equation to solve the electrostatic potential and charge distribution, while DLVO takes into account both the electrostatic and van der Waals (vdW) interactions to further quantify the interaction energy and surface adsorption effects. Despite their broad applications, these models do not reveal the key interfacial liquid structure. In fact, both GCS and DLVO theory treat solvents as uniform dielectric backgrounds, and the obtained charge and potential profiles are always smooth with no oscillations. Certain macroscopic properties, such as double layer capacitance and colloidal stability in dilute aqueous solutions, can be approximately predicted via these "smoothened" models.[27,71] However, caution must be taken to avoid overusing them for problems where the molecular-level interfacial structure plays a key role, such as in electrocatalysis,[11] interfacial passivation in batteries,[72] and capacitive charging in highly ionic electrolytes.[29]

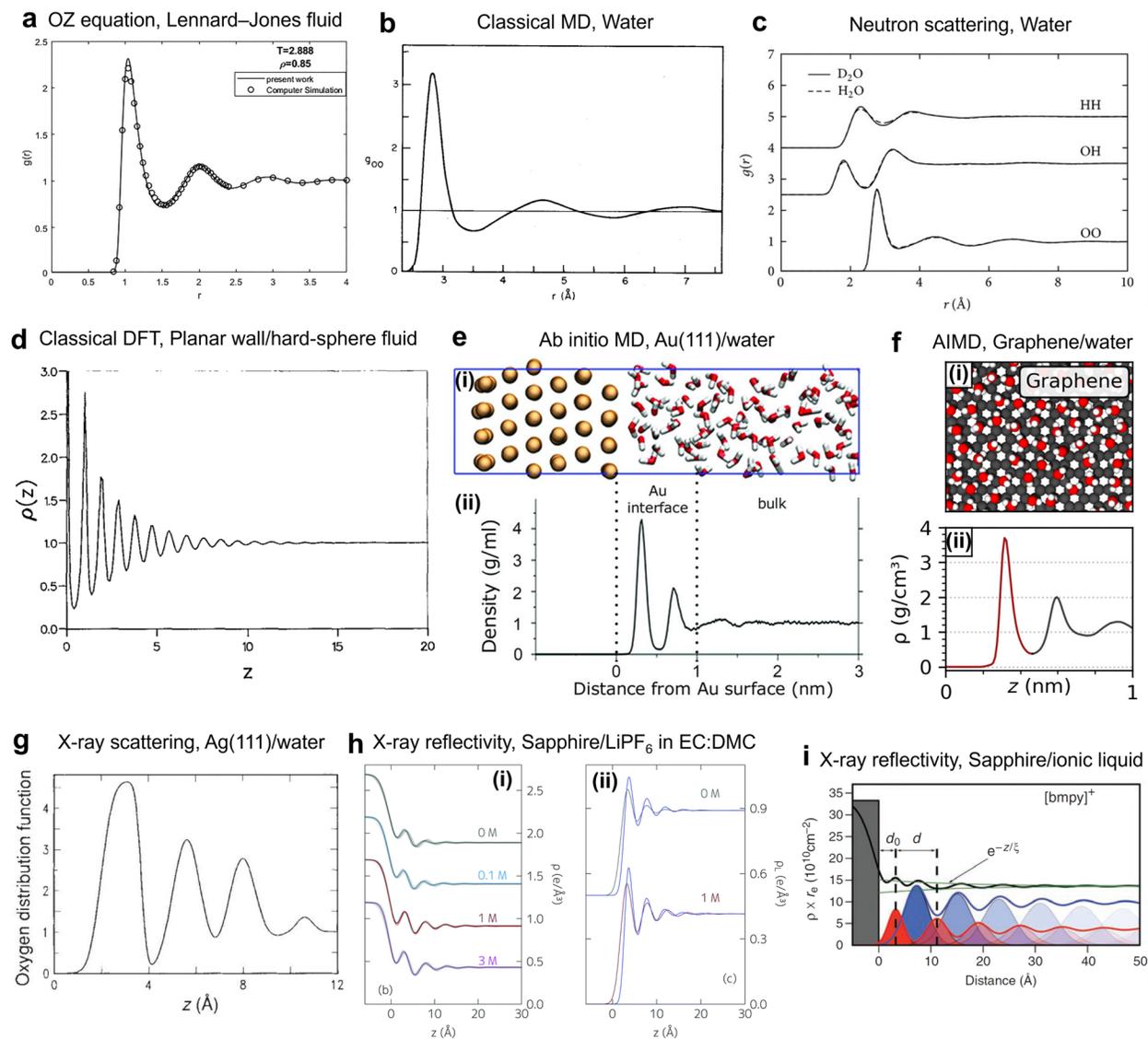

**Figure 4. Gallery of existing results on bulk and interfacial liquid structure from computational and scattering methods.** (a) PCF derived from OZ equation of a general bulk Lennard-Jones fluid.[73] Reproduced with permission from ref [73]. Copyright 2021 Taylor & Francis. (b) Classical MD simulation of the O–O PCF for bulk water.[74] Reproduced with permission from ref [74]. Copyright 1974 AIP Publishing. (c) PCF of bulk $H_2O$ and $D_2O$ derived from neutron scattering measurements.[75] Reproduced from ref [75]. Copyright 2013 The Author. (d) Classical DFT simulation of liquid density vs z profile for a general hard-sphere fluid next to a planar hard wall.[76] Reproduced with permission from ref [76]. Copyright 1992 AIP Publishing. (e) AIMD simulation of

Au(111)/water interface, including d(i) snapshot of the atomic structure and d(ii) mass density of the water vs distance from the Au surface.[23] Reproduced with permission from ref [23]. Copyright 2014 The American Association for the Advancement of Science. (f) AIMD simulation of graphene/water interface, including f(i) snapshot of the top-down view of the first water layer on graphene and f(ii) water density vs z.[77] Reproduced from ref [77]. Copyright 2024 The Authors. (g) Oxygen distribution vs z of 0.1 M NaF solution in water next to Ag(111) at 0.52 V vs the potential of zero charge, extracted from X-ray scattering measurements.[33] Reproduced with permission from ref [33]. Copyright 1994 Springer Nature. (h) Electron density profile derived from X-ray reflectivity measurements for sapphire (001)/LiPF$_6$ in EC:DMC (weight-ratio 1:1) interface, including h(i) the total electron density of the solid and liquid and h(ii) the electron density of only the liquid.[34] Reproduced with permission from ref [34]. Copyright 2018 Royal Society of Chemistry. (i) Cation (red), anion (blue), and total (black) electron densities derived from X-ray reflectivity measurements of an ionic liquid, [bmpy]$^+$[FAP]$^-$ (1-butyl-1-methylpyrrolidinium tris(pentafluoroethyl)trifluorophosphate) next to sapphire (0001).[35] Reproduced with permission from ref [35]. Copyright 2008 The American Association for the Advancement of Science.

A better starting point is the statistical mechanics description of bulk liquid structure. Although bulk liquid has a uniform time-averaged density, it is not structureless. Rather, the key structure can be described by the pair correlation function (PCF), which reveals the density distribution as a function of displacement from a reference molecule/particle. The PCF of most known liquids, if not all, has a damped oscillation profile, as confirmed by atomistic simulations and X-ray and neutron scattering measurements (Figure 4a–c).[73–75,78,79] For simple hard sphere fluids, the classical Ornstein-Zernike (OZ) equation also predicts oscillatory decays as long as the liquid density is above a threshold (the Fisher-Widom line).[80,81] The universality of the oscillatory decay profiles in PCF among a large range of different liquids confirms that this feature is due to the finite size effects of the solvent/solute molecules, rather than any specific molecular interactions.

At solid–liquid interfaces, when the solid surface is flat (either a theoretical wall or an ideal single crystal consisting of individual atoms), classical density functional theory (DFT), classical MD, and ab initio MD (AIMD) all predict oscillatory decay profiles of the liquid density vs z (Figure 4d–f).[23,38,49,76,77,81,82] Experimental measurements using X-ray/neutron scattering also verified the oscillatory density profiles for water/aqueous solutions, organic solvents/electrolytes and ionic liquids at solid surfaces (Figure 4g–i).[33–35] Therefore, it is safe to conclude that density oscillations are universal for most, if not all, liquid systems adjacent to solid surfaces. In addition, it has also been commonly observed that the oscillation period of the interfacial liquid density vs z is the same as or similar to that of the PCF of the corresponding bulk liquid,[23,33–35,38,49,76,77,81,82] revealing that these oscillations share the same origin—the finite size of the liquid molecules/ions.

**4. Solvent Tip Approximation to Connect 3D-AFM Force to Intrinsic Liquid Density**

To quantitatively interpret the 3D-AFM force maps, we first need to fully appreciate the dynamic nature of the interfacial liquids. Although the liquid molecules are structured at the interface, they are continuously moving around, and it is the time-averaged density that exhibits oscillation patterns. Therefore, the traditional perspective of AFM probe interacting with fixed or deformable solids is no longer valid. While the long-range interactions (e.g. vdW force) between the probe and substrate can lead to a monotonic background force, the oscillatory part of the force, $\Delta f$, should be directly related to the configurational free energy of the interfacial liquid molecules. That is, as the probe penetrates through the interfacial liquid, the molecules are forced to dynamically rearrange, leading to a change in the configurational partition function and thus a different free energy. The spatial gradient of free energy gives rise to the oscillatory force $\Delta f$.

**4.1. Key hypotheses and derivation of the STA model.** A simple yet powerful analytical model that relates AFM force to liquid density is STA, initially proposed by Watkins and Reischl[53] and Amano et al.[54] As illustrated in Figure 5, a key hypothesis of the STA model is that a solvent molecule (same as the solvent in the surrounding liquid) is "fixed" right below the very end of the probe. This terminating molecule is not intentionally or manually attached to the probe. Rather, it sticks around the tip due to a few possible reasons: 1) chemisorption; 2) physisorption; 3) it corresponds to the terminating point of the first solvation layer of the probe. This terminating molecule can be either strictly attached to the probe (e.g., chemisorption) or dynamically moving around yet are next to the end of the probe most of the time (e.g., the first solvation layer). In AC mode 3D-AFM imaging, the probe oscillation amplitude is typically no more than 1–2 Å. With such small dynamic motion, the terminating molecule is the "hot spot" that mainly determines the probe–sample interaction force. For simplicity, the whole probe is treated as this terminating molecule in the STA model.

When the probe is far away from the solid surface (Figure 5b), the interfacial liquid is in the intrinsic, unperturbed state with full configurational partition function. In this state, the system's Helmholtz free energy is:

$$F = -k_B T ln Q = -k_B T ln Z_N + const, \qquad (1)$$

where $Q$ is the total partition function, while $Z_N$ is the configurational partition function. When the probe is moved close to the surface, at a position $\boldsymbol{r}$, the terminating molecule poses constraints to the interfacial liquid molecules' positional configurations. The perturbed configurational partition function can be written as $Z(\boldsymbol{r})$. As the probe move from far away (bulk liquid) to position $\boldsymbol{r}$ (near surface), the change in free energy is:

$$\Delta F(\boldsymbol{r}) = -k_B T ln\left(\frac{Z(\boldsymbol{r})}{Z_N}\right). \qquad (2)$$

Watkins and Reischl hypothesized that the time-averaged interfacial liquid density $\rho(\boldsymbol{r})$ follows a simple relation:[53]

$$\frac{\rho(\boldsymbol{r})}{\rho_0} = \frac{Z(\boldsymbol{r})}{Z_N}, \qquad (3)$$

where $\rho_0$ is the bulk liquid density. Therefore, the probe–sample interaction force can be obtained as:

$$\Delta f(z) = -\frac{dF(z)}{dz} = \frac{k_B T}{\rho(z)} \frac{d\rho}{dz}. \qquad (4)$$

Eq. 4 is the well-known STA model.

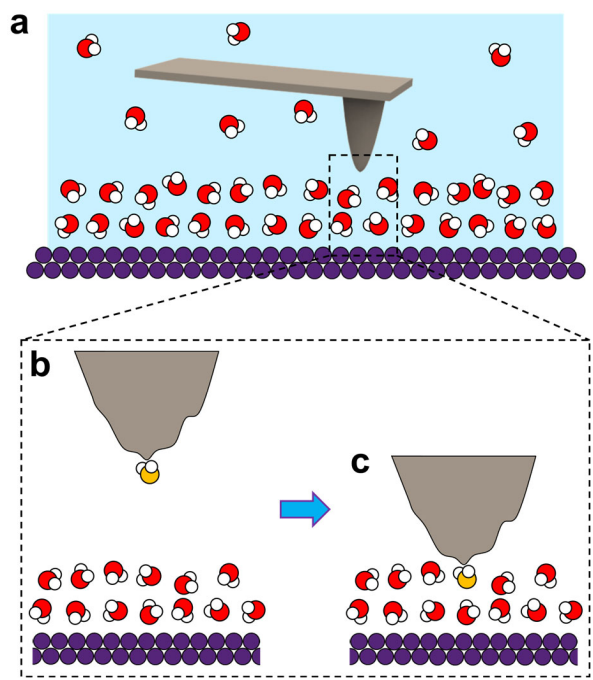

**Figure 5. Schematic of the STA model.** (a) Overall configuration of AFM probe immersed in liquid near a solid surface. (b) Expanded view of the end of the probe with a terminating molecule in bulk liquid. (c) Probe with a terminating molecule at the interfacial liquid region.

Despite the simplicity, STA captures the key mechanism of probe–sample interaction. In contrast to pure solid samples where probe–substrate interaction results from the modulation of the system's enthalpy or internal energy, interaction with interfacial liquids is manifested as both enthalpic and entropic effects. For simple hard-sphere liquids where long-range interactions are absent, it is expected that entropic effects play a dominant role in determining both the interfacial density oscillations and the resulting probe–liquid interaction force. Even though the presence of the probe unavoidably perturbs the intrinsic interfacial liquid configuration, such perturbation can be quantified and used to back-calculate the original, unperturbed interfacial liquid density profile. STA model provides a simple analytical formula for such back-calculation.

**4.2. Comparison of STA model with experimental results.** STA model has been widely used in quantitative interpretation of 3D-AFM force curves and maps, and has been shown to be consistent with experimental results and with atomistic simulations.[43,44,66,83–85] Figure 6 summarizes two representative work where quantitative comparison was presented. Fukuma lab used FM 3D-AFM and MD simulations to investigate fluorite (111)/water interfaces.[66,84] From MD simulated interfacial water density, they used STA to extract the theoretical force map (Figure 6a(i)) and compared them with the experimental force map (Figure 6b(i)). The key features in these maps are highly consistent. Further comparison of the experimental and STA force curves at different atomic spots also reveals strong agreements (Figure 6a(ii) and b(ii)), where the experimental and STA force values are within a factor of two from each other. Another work from Kühnle lab also observed rough quantitative agreement between the STA force curve and realistic probe–sample force curve for calcite/water interface (Figure 6c).[43]

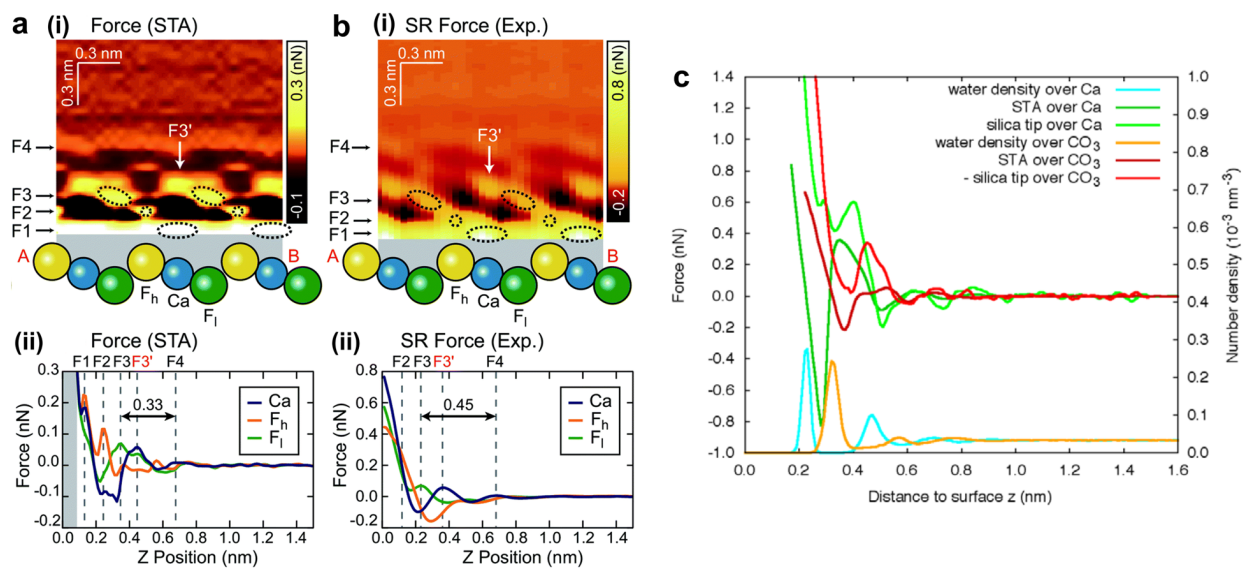

**Figure 6. Comparison of STA model with realistic results.** STA force map (a(i)) and curve (a(ii)) compared to experimental force map (b(i)) and curve (b(ii)) for fluorite (111)/water interface.[66] Reproduced from ref [66]. Copyright 2016 The Authors. (c) Comparison of STA force curve and the force curve of a realistic silica tip at different sites of calcite/water interface.[43] Reproduced with permission from ref [43]. Copyright 2018 American Physical Society.

Considering that the STA model treats the probe as a single molecule, its quantitative agreement with experimental/realistic data is remarkable. Such consistency confirms that the "hot spot" effect is indeed reasonable. That is, although the whole AFM probe is large (e.g., at least a few microns tall) and the tip radius can be at least a few nanometers, the critical region sensing the interfacial liquids is at the very end of the probe, approximated as one terminating molecule.

## 5. Explicit Tip Models for Understanding 3D-AFM Maps

**5.1. MD simulations.** To improve the precision of interfacial liquid density quantification, explicit tip models have been developed, where the tip is approximated not as a single molecule, but a nanocluster consisting of individual atoms.[40,45,48,85–89] Through MD simulations of a series of systems with different probe–substrate distance, the force-distance curves are extracted. These all-atom models have successfully reproduced the density and force oscillations and achieved qualitative agreement with experimental force curves/maps. Nevertheless, compared to the simple STA model, the all-atom MD results have not exhibited obvious improvements. The MD simulated force values still tend to deviate from the experimental results within a factor of 2–3.

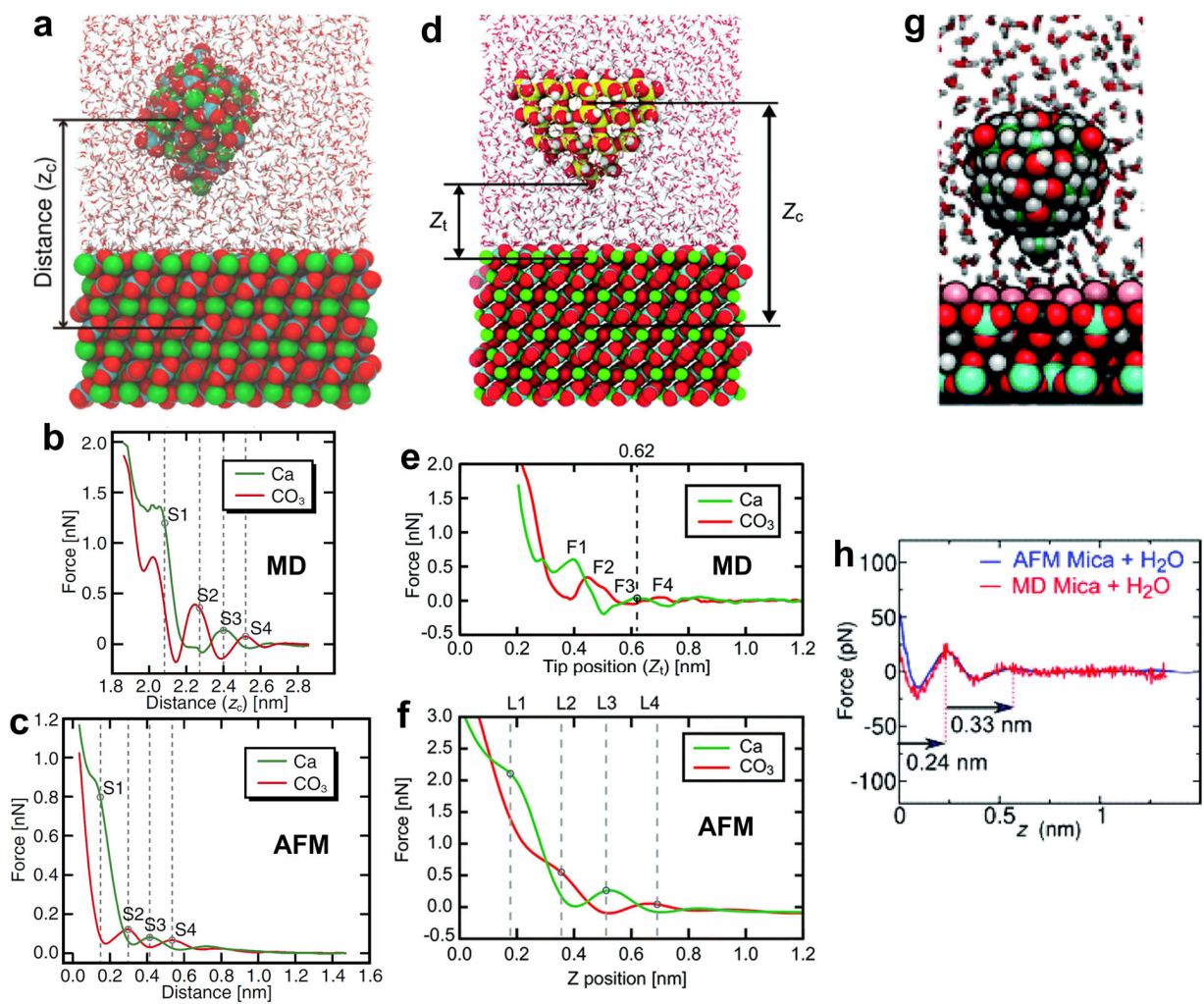

**Figure 7. All-atom MD simulation of the 3D-AFM imaging process.** (a) Snapshot of a model AFM tip (calcite cluster) near a calcite/water interface, (b) MD simulated force curves at the Ca and $CO_3$ sites, and (c) experimental AFM force curves at the corresponding sites (using a diamond-like carbon tip).[45] Reproduced from ref [45]. Copyright 2015 The Authors. (d–f) Another work showing a simulated OH-terminated silica tip near a calcite/water interface, the simulated force curves, and the experimental force curves (using a silicon tip with native oxide).[85] Reproduced from ref [85]. Copyright 2020 The Authors. (g, h) Snapshot of a simulated AFM tip (hydroxylated diamondoid cluster) near a mica/water interface, and the experimental (using a silicon tip with native oxide) and simulated force curves at the mica/water interface.[40] Reproduced from ref [40]. Copyright 2021 The Authors.

The first comprehensive MD simulation of the 3D-AFM imaging process is by Fukuma, Foster, and co-workers, with key results summarized in Figure 7a–c.[45] Experimentally, they measured calcite/water interface using an AFM tip composed of diamond-like carbon. In the MD simulation, however, a calcite cluster was used to function as the tip (Figure 7a), as the authors hypothesized that calcite may have adsorbed onto the AFM tip during the imaging process. The simulated interfacial water density and tip–sample interaction force both exhibited damped oscillations, as expected. Compared to the experimental results, the force oscillation features are overall similar, although the values tend to deviate by a factor of ~2–3. A later work by Fukuma explored multiple different tip structures both in simulation and in experiments for calcite/water interfaces, and the results show the same qualitative similarities in the oscillation features yet quantitative differences in force values (Figure 7d–f).[85] A recent work by Comer and Garcia achieved quantitative agreements in the simulated and experimental force curves for mica/water interface, where they used silicon tips (with native oxide) in experiments and hydroxylated diamondoid cluster as the tip in simulation (Figure 7g, h).[40] Considering the obvious differences in the experimental and simulated AFM tip, the quantitative agreements in the obtained force values are either due to coincidence or indicate that the interaction force is insensitive to the exact tip composition (similar to the assumptions in the STA model).

The comparable accuracy of the all-atom MD simulation and the simple STA model has profound implications. In fact, the experimental results themselves tend to have variations in the force values with a factor of 2–3, even when the same type of AFM tip was used in separate measurements of the same samples.[85] Such fluctuations are likely due to the random variations in the exact chemical and structural details at the very end of the tip, which to date still cannot be accurately predicted by either the MD or STA models. Consequently, despite the significantly higher computational cost, MD simulations incorporating the all-atom tip structure have not been offering higher accuracy than the STA model.

**5.2. Classical DFT.** In contrast to the all-atom MD and single-molecule STA models, classical DFT provides an intermediate theoretical framework, treating the AFM tip as rigid, closely packed hard spheres while modeling the liquid as mobile hard spheres. Seminal works by Tarazona and Garcia laid the foundation for DFT analysis of 3D-AFM imaging processes.[38,90] According to this theory, the oscillatory tip–sample interaction force is due to entropic effects associated with the packing of interfacial liquid molecules. Similar to the STA model, classical DFT analysis also leads to the conclusion that the AFM force, resulting from the perturbation of the configurational free energy of the interfacial liquid, is directly related to the intrinsic, unperturbed liquid density. In fact, classical DFT further predicts the approximate form of both the interfacial liquid density and the oscillatory force:[38,90]

$$\Delta\rho(z) = \rho(z) - \rho_0 \approx \Delta_\rho e^{-\alpha z} \cos(qz - \varphi_\rho), \tag{5}$$

$$\Delta f(z) \approx \Delta_f e^{-\alpha z} \cos(qz - \varphi_z), \tag{6}$$

where $\Delta_\rho$ and $\Delta_f$ are the amplitude factors, $1/\alpha$ is the common decay length for both the density and force, $2\pi/q$ is the common oscillation period, and $\varphi_\rho$ and $\varphi_z$ are the phase factors. $\alpha$ and $q$ are factors that characterize the bulk liquid structure, while the amplitude and phase terms ($\Delta_\rho$, $\Delta_f$, $\varphi_\rho$, $\varphi_z$) contain information on the interfacial liquid and tip–liquid interaction.

As shown in Figure 8a, when the tip approaches the substrate surface, the solvation layers near the tip interfere with those next to the substrate, thus modulating the liquid's configurational entropy. Such entropic modulations result in the change of the overall free energy, and the gradient of the free energy is the force. The intrinsic, nonperturbed liquid density profile determines the pattern of the entropic modulation and thus the eventual force oscillation profile sensed by the probe. Since the intrinsic liquid density has the form of damped oscillation as a generic DFT result, the force also exhibits a similar pattern with a phase shift (Figure 8b).

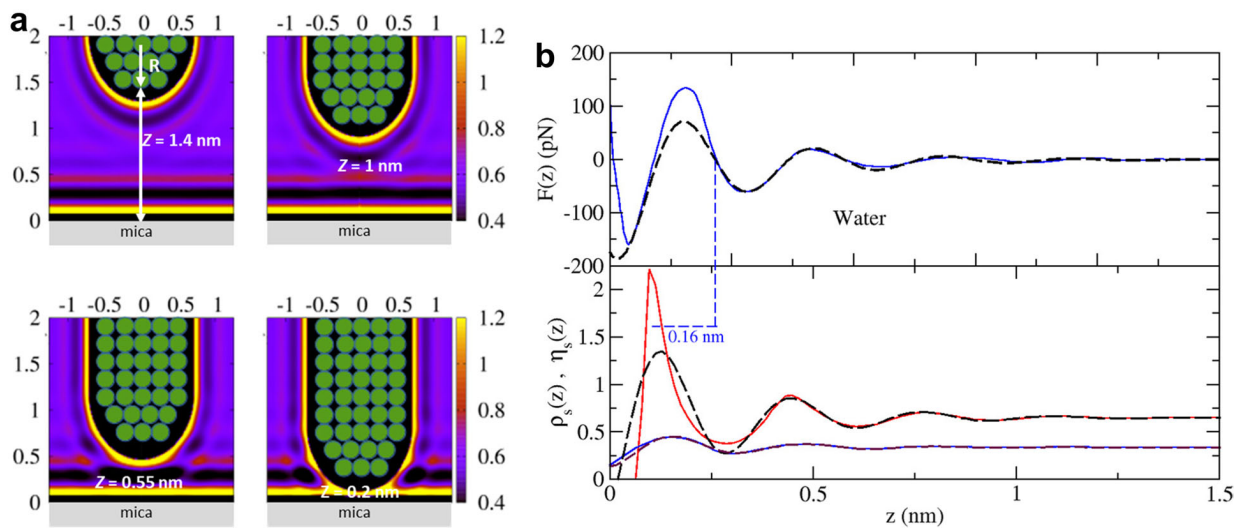

**Figure 8. Classical DFT analysis of the 3D-AFM imaging process.**[38] (a) The simulated water density distribution in the presence of the tip (close packed spheres) and substrate (mica), at four different tip–substrate separations. (b) (top) Experimental AFM force curve at mica/water interface (blue line) and its analytical fit (black dashed line) and (bottom) DFT results for the liquid density (red line) and its analytical fit (black dashed line) in the absence of the AFM tip. Reproduced with permission from ref [38]. Copyright 2020 American Chemical Society.

The DFT results offer a powerful framework for in-depth analysis of the 3D-AFM results. Eqs. 5 and 6 are great starting points for extracting the key structural parameters of both the bulk and interfacial liquids. One limitation, however, is that the first force/density peak for realistic liquids tends to deviate from these analytical formulas, as evident in the example in Figure 8b. This is because the hard sphere model does not take into account the specific interactions between the liquid molecules and the substrate, which gives rise to extra enthalpy or internal energy terms that strongly modulate the first solvation peak.

## 6. Outlook

After more than one decade of development, 3D-AFM has reached a stage where the imaging protocols have been well established yet the data interpretation is still elusive. The key open question is the precise, quantitative connection between the experimental force map and the interfacial liquid density distribution. Existing experimental, theoretical, and computational studies have provided comprehensive evidence that 3D-AFM maps do reveal the intrinsic, unperturbed liquid density profiles. While the precise force–density correlation has not been possible, the simple STA model can already perform a rough density–force conversion, with quantitative values accurate within a factor of 2–3.

A natural question that arises is: since the AFM tip has a typical radius of at least a few nanometers, how can a single molecule be used to approximate the tip–sample interaction? The underlying mechanism is similar to the well-known atomic contrast principles in scanning tunneling microscopy (STM) and ultrahigh vacuum non-contact AFM (NC-AFM).[91,92] In all of these atomic-resolution scanning probe imaging methods, the tip radius is rarely smaller than a few nanometers or even 10s of nanometers. However, despite the large tip radius, the very end atom (or few atoms) is dominantly responsible for the imaging process, modulating either the electron tunneling rate or the tip–sample interaction force. For the specific case of 3D-AFM at solid–liquid interfaces, studies by Onishi[93] and Fukuma[94] have both observed that the spatial resolution and interaction force is independent of the tip radius within the range of ~10 nm to 100s of nm. Key results by Onishi and co-workers are summarized in Figure 9, where two different tips, with radii of 10 nm and 250 nm respectively, produced nearly identical atomic resolution and force-distance curves. The results were attributed to the "minitip" effect, where the very end of the probe features a local protrusion that acts as a minitip. The minitip effect is unintentional yet unavoidable, resulting from random variations in the local structures of a microfabricated probe. Such a minitip can be as small as one or a few atoms, and the surrounding first solvation layer can be approximated as a single molecule, thereby rationalizing the STA model.

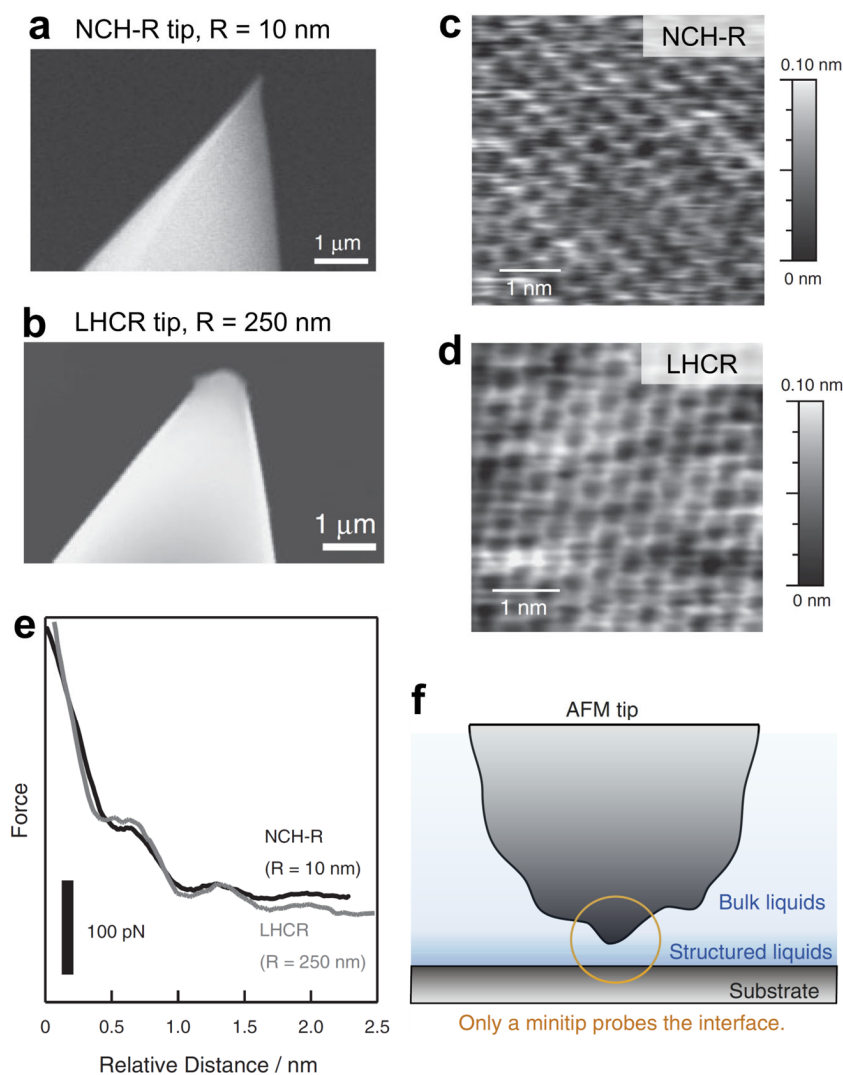

**Figure 9. Comparison of 3D-AFM imaging using two different tips.**[93] Scanning electron microscopy images of two different probes, (a) NCH-R with a radius of 10 nm and (b) LHCR with a radius of 250 nm. Both probes consist of Si with native oxide. Surface topography images of mica in 1 M KCl aqueous solution using (c) NCH-R probe and (d) LHCR probe. (e) Force-distance curves obtained with NCH-R (black) and LHCR (gray) probes at a dodecanethiol self-assembled monolayer/hexadecane interface. (f) Schematic of the minitip imaging interfacial liquids. Reproduced from ref [93]. Copyright 2012 The Authors.

Although the minitip imaging effect is similar among 3D-AFM in liquid and NC-AFM in vacuum, the origin of the measured force is distinct. In NC-AFM, the force arises from the direct interaction energies—such as vdW, chemical, and electrostatic—between the terminal atom(s) of the tip and the solid surface. Entropic effect is mostly negligible, especially at cryogenic temperatures. For 3D-AFM in liquid, the tip–substrate interaction energy still contributes to the force, but mainly to the monotonic background. The oscillatory force $\Delta f$ originates from the tip-modulation of the configurational free energy of the interfacial liquid, which is likely mainly entropic in nature, with negligible contribution from the direct tip–molecule or tip–substrate interaction energy/enthalpy.

The existing theoretical/computational studies of the 3D-AFM imaging process have mainly focused on single-component liquid systems. Nevertheless, the key conclusions can be easily generalized to multi-component liquid electrolyte systems relevant to practical applications. Experimentally, for dilute aqueous solutions with salt concentration less than ~1 M, 3D-AFM force maps of the interfacial liquid are nearly the same as those in pure water.[39,43,45,46,48,65] The only reliably observed difference is the specific adsorption of certain ionic species, such as $K^+$ cations on mica.[39] These results align with classical liquid theory, which predicts that the PCF of binary hard sphere mixture fluids closely resembles that of the predominant hard sphere component.[95–97] Since water is the major component in the dilute aqueous solutions, it is not surprising that the interfacial density and force distribution are nearly independent of the salt species and concentration. However, when the electrolyte is highly concentrated or even purely ionic (i.e., ionic liquid), the interfacial structure becomes quite complicated and is likely beyond the scope of the simple single-component models.[41,47,49–51]

In summary, we have surveyed the 3D-AFM imaging mechanism and key results on solid–liquid interfaces, the intrinsic interfacial liquid density profiles, and the theoretical and computational approaches to connect the 3D-AFM force maps to the liquid density distributions. For one-component solvents and dilute electrolyte solutions, STA model is largely sufficient for an approximate density–force correlation. Further investigations may focus on: 1) optimizing the 3D-AFM tip preparation and imaging protocols, so that the tip's surface termination is more well-defined and the force maps have minimum tip-to-tip variations; 2) improving the accuracy of the STA model or other analytical/computational methods, with the goal of more precisely converting force to liquid density; and 3) developing models to achieve density–force correlation for more complex liquid systems such as ionic liquids and highly concentrated electrolytes.

**Associated Content**

**Notes**

The authors declare no competing financial interest.

## Biographies

**Qian Ai** earned his bachelor's degree in Materials Science and Engineering from Southern University of Science and Technology in 2021. He is currently a Ph.D. student in the Department of Materials Science and Engineering at University of Illinois Urbana-Champaign. His research is focused on imaging and understanding the structure of interfacial liquids for renewable energy systems.

**Lalith Krishna Samanth Bonagiri** earned his bachelor's degree in Mechanical Engineering from the Indian Institute of Technology Madras in 2019. He is currently pursuing a Ph.D. in the Department of Mechanical Science and Engineering at the University of Illinois Urbana-Champaign. His research topic is on advancing atomic force microscopy (AFM) for imaging electrical double layers. Additionally, he is integrating deep learning techniques to enhance AFM imaging capabilities.

**Amir Farokh Payam** is an associate professor (senior lecturer) at Ulster university. His research focuses on theoretical and experimental development of atomic force microscopy systems, nano/micro-resonators, solid-liquid and materials characterization and MEMS/NEMS.

**Narayana Aluru** is a professor in the Department of Mechanical Engineering and a core faculty in the Oden Institute for Computational Engineering and Sciences at the University of Texas at Austin. His research focuses on the development of multiscale methods and using them to probe physics of solid-liquid interfaces, nanofluidics, and nanomaterials.

**Yingjie Zhang** is an assistant professor at the University of Illinois Urbana-Champaign. His research focuses on in situ characterization of solid-liquid interfaces, electrocatalysis, and chemical imaging of biological cells.

## Acknowledgments

Q.A., L.K.S.B. and Y.Z. acknowledge support from the National Science Foundation under Grant No. 2137147. A.F.P. acknowledges the support by the Department for Economy, Northern Ireland through US-Ireland R&D partnership grant No. USI 186. N.R.A. acknowledges support from the National Science Foundation under Grant No. 2137157.